\documentclass[preprint,nofootinbib,aps,11pt]{revtex4-1}
\usepackage{graphicx}% Include figure files
\usepackage{dcolumn}% Align table columns on decimal point
\usepackage{bm}% bold math
\usepackage{epsfig,amsmath}
\usepackage{amssymb}
\usepackage{subfigure,esint}
\usepackage{float}
\usepackage[usenames,dvipsnames]{color}
\usepackage[pagebackref=false, colorlinks=true]{hyperref}
\definecolor{redish}{rgb}{0.7,0.2,0.0}  % color defined in (r=red,g=green,b=blue) model
\definecolor{bluish}{rgb}{0.2,0.5,0.8}

\hypersetup{linkcolor=redish,          % color of internal links
                  citecolor=blue,        % color of links to bibliography
                  filecolor=magenta,      % color of file links
                  urlcolor=blue}          % color of the url

\DeclareFontFamily{U}{rsfs}{}         % Formal Script            %
\DeclareFontShape{U}{rsfs}{m}{n}{<5> rsfs5 <6><7> rsfs7          %
  <8><9><10><10.95><12><14.4><17.28><20.74><24.88> rsfs10}{}     %
\DeclareMathAlphabet{\mathfs}{U}{rsfs}{m}{n}

\begin{document}
\author{Rajibul Shaikh$^{\;1,}$}
\email{rshaikh@iitk.ac.in}
\author{Pankaj S. Joshi$^{\;2,}$}
\email{psjprovost@charusat.ac.in}
\affiliation{${}^{1}$ Department of Physics, Indian Institute of Technology, Kanpur 208016, India}
\affiliation{${}^{2}$ International Center for Cosmology, Charusat University, Anand 388421, Gujarat, India}

\title{\Large Can we distinguish black holes from naked singularities by the images of their accretion disks?}

\bigskip

\begin{abstract}
We study here images of thin accretion disks around black holes and two classes of naked singularity spacetimes and compare these scenarios. The naked singularity models which have photon spheres have single accretion disk with its inner edge lying outside the photon sphere. The images and shadows created by these  
models mimic those of black holes. It follows, therefore, that further and more detailed analysis of the images and shadows structure in such case is needed to confirm or otherwise the existence of an event horizon for the compact objects such as the galactic centers. However, naked singularity models which do not have any photon spheres can have either double disks or a single disk extending up to the singularity. The images obtained from such models significantly differ from those of black holes. Moreover, the images of the two classes of naked singularities in this latter case, differ also from one another, thereby allowing them to be distinguish from one another through the observation of the images.
\end{abstract}
\maketitle

%%%%%%%%%%%%%%%%%%%%%%%%%%%%%%%%%%%%%%%%%%%%%%%%%%

\section{Introduction}

The recent observations of the images of the galactic center for the galaxy M87 by the event horizon telescope
\citep{EHT1,EHT2,EHT3}
have been a matter of great interest and scientific significance.
That is because, these could possibly reveal a great deal of information on the nature of the central object, its dynamical behavior, and especially the existence
or otherwise of the event horizon in the galactic center as 
we continue to refine our observations in the coming time as the 
EHT goes to the next phases of its observations.

Basically, what the EHT has reported and observed so far is an image containing a shadow (a darker region over a brighter background) of the central supermassive object of M87 in some detail. Such an image can be created if the central object is a black hole, which means it has necessarily an event horizon, and also therefore a photon ring. The event horizon always implies that there must be a photon sphere surrounding the central object, which in turn is a sufficient condition for creating the shadow image. As we have shown recently in \cite{shaikh_2018a}, even if the event horizon covering the central ultra-dense object or the central singularity does not exist, such a naked singularity also can have a photon sphere around the same for a range of allowed physical parameter values. This again creates a shadow of the central object. Therefore, the existence of a shadow image is not sufficient to define and determine fully the nature of the central object, which could be a black hole or a naked singularity or a horizonless other ultracompact object, that developed during the gravitational collapse that asymptotically produced these final states within the galactic center. This issue of whether a shadow implies an event horizon has also been addressed in \cite{shadow_horizon,abdikamalov_2019}. Also, see \cite{broderick_2006b,bambi_2009,bambi_2019d, bambi_2013a,nedkova_2013,ohgami_2015,mustafa_2015, abdujabbarov_2016b,rajibul_2018b,gyulchev_2018,amir_2018, rajibul_2018c,rajibul_2019,vincent_2016} for some works on shadow cast by various horizonless objects.

Typically, black hole-like compact objects in the universe are formed as the end state of gravitational collapse of matter. A necessary implication of general relativity is that continual gravitational collapses always lead to spacetime singularities. According to cosmic censorship hypothesis, such a singularity must always be hidden within an event horizon, thereby allowing no signals from the vicinity of the singularity to be seen by distant observers. However, a conclusive proof of the hypothesis at present is lacking. On the other hand, in many studies of gravitational collapse, it turns out that the end state of collapse is not necessarily always a black hole \citep{christodoulou_1984,ori_1987,choptuik_1993,joshi_book, joshi_2011} (see also \citep{joshi_1993,singh_1996,jhingan_1996,mena_2000}); a naked singularity can also form as the end state of continual collapse, depending on the regular initial conditions. Recently, it has been shown that a collapsing matter cloud with physically suitable initial conditions can equilibrate itself to form a static naked singularity in asymptotic time \citep{JMN1,JMN2,banik_2017}.

It is therefore very much important to study various aspects of naked singularities which can help us testing their existence in astrophysical scenarios. In light of this, there has been growing interest in the past decades to explore various aspects of naked singularity such as strong gravitational lensing \citep{lensing1,lensing2,lensing3,lensing4,lensing5,lensing6,lensing7,  lensing8,lensing9}, images and shadows \citep{image1,image2,image3,image4}, perihelion precession of timelike particle orbits \citep{perihelion1, perihelion2,perihelion3,perihelion4,perihelion5} etc. In \cite{shaikh_2018a}, we considered a simplified model of spherical accretion onto the central object and studied shadow and images of Joshi-Malafarina-Narayan (JMN) naked singularities obtained in \cite{JMN1,JMN2}. We use the terms JMN-1 and JMN-2, respectively, for the naked singularities of \cite{JMN1} and \cite{JMN2}. One could ask, whether or not qualitatively similar conclusions obtained in \cite{shaikh_2018a} for spherical accretion models will arise as for the shadow and images are concerned, when photons from the particles going around in a thin accretion disk are considered. In this work, therefore, we consider thin accretion disks around JMN-1 and Janis-Newman-Winicour (JNW) \citep{JNW} naked singularity and study their images. Various aspects of the JMN-1 naked singularity have been studied in \cite{shaikh_2018a,JMN2,lensing4,perihelion4}. The JNW naked singularity spacetime has also attracted much attention from various perspectives \citep{lensing1,lensing2,lensing3,lensing7, perihelion5,chowdhury_2012,gyulchev_2019}. Timelike circular geodesics and accretion disk in JNW naked singularity was studied in \cite{chowdhury_2012}. The images of JNW singularity with thin accretion disk for the case when there is a photon sphere was considered in \cite{gyulchev_2019}. This latter case qualitatively mimic that of black hole. In this work, along with the JMN-1 naked singularity, we consider the other case of the JNW singularity (i.e., the case when there is no photon sphere) and study their images. Thin accretion disk and their various aspects in JMN-2 \citep{JMN2}, Reissner-Nordstrom \citep{pugliese_2011,schee_2015}, rotating JNW \citep{kovacs_2010}, Kerr-Taub-NUT \citep{chakraborty_2019} and Kerr naked singularity \citep{stuchlik_2010,stuchlik_2012} has also been considered. See \cite{luminet_1979,takahashi_2004,bambi_2012,tian_2019} for some works on images of black holes with thin accretion disks.

The essential conclusion that follows from our study here is, even in the case of the thin accretion disk forming the images, rather than the simplified case of radial spherical accretion considered earlier, the naked singularities having photon spheres do create images and shadow, similar in nature to the black hole case. The key indication that follows is that, to decide on the presence or absence of an event horizon, more detailed and further analysis will be needed, and mere existence of a shadow does not by itself imply the existence of black hole or event horizon. On the other hand, when there are no photon spheres, the images of naked singularities significantly differ from those of black holes. In this latter case, one can clearly distinguish a naked singularity from a black hole.

The plan of the paper is as follows. In Sec. \ref{sec:spacetimes}, we briefly summarize the naked singularity spacetimes under consideration. We consider Novikov-Thorne thin disk model \citep{novikov_1973,page_1974} around the naked singularities and study the radiation emitted by the disks in Sec. \ref{sec:disks}. In Sec. \ref{sec:tracing}, we discuss the technique for tracing (backward in time) the photons emitted from the disks and received by a distant observer. In Sec. \ref{sec:images}, we produce the images by using continuous color maps describing the redshifted apparent radiation flux received by a distant observer and compare among different scenarios. Finally, we conclude in Sec. \ref{sec:conclusions} with some discussions on our results.

\section{The black hole and naked singularity spacetimes}
\label{sec:spacetimes}

In this work, we consider thin accretion disks around Schwarzschild black hole and two classes of naked singularity and produce their images. The first class of naked singularity we consider is the JMN-1 spacetime which is described by the following metric \citep{JMN1},
\begin{eqnarray}
ds^2 &=& -(1-M_0)\left(\frac{r}{R_b}\right)^{M_0/(1-M_0)}
dt^2+\frac{dr^2}{1-M_0} +r^2\left(d\theta^2+\sin^2\theta
d\phi^2\right),
\label{eq:JMN1}
\end{eqnarray} 
where the parameter $M_0$ is limited to the range $0 \leq M_0 \leq 4/5$ (the upper limit corresponds to the requirement that the sound speed should not exceed unity). The spacetime contains a time-like naked singularity at $r=0$ and is matched at its outer radius $r=R_b$ to the Schwarzschild geometry,
\begin{equation}
ds^2=-\left(1-\frac{2M}{r}\right)
dt^2+\frac{dr^2}{1-\frac{2M}{r}}+r^2\left(d\theta^2+\sin^2\theta
d\phi^2\right),
\end{equation}
where the total mass $M$ is related to $M_0$ and $R_b$ by
\begin{equation}
M = \frac{1}{2}M_0 R_b.
\label{eq:MRb}
\end{equation}
The JMN-1 naked singularity have a photon sphere when $M_0\geq 2/3$, or equivalently when $R_b\leq 3M$. In this case, the photon sphere lies in the exterior Schwarzschild geometry \citep{shaikh_2018a}. It does not have any photon sphere when $M_0<2/3$, or equivalently when $R_b>3M$. Therefore, models with $M_0\geq 2/3$ do produce shadows similar to those of black holes. On the other hand models with $M_0<2/3$ do not produce any shadow. Instead, they produce an interesting full-moon image. See \cite{shaikh_2018a} for more details.

The second class of naked singularity we consider is the JNW singularity which is the solution of the Einstein-scalar field equation. The spacetime geometry is given by \citep{JNW}
\begin{equation}
ds^2=-\left(1-\frac{2M}{\gamma r}\right)^\gamma dt^2+\frac{dr^2}{\left(1-\frac{2M}{\gamma r}\right)^\gamma}+r^2\left(1-\frac{2M}{\gamma r}\right)^{1-\gamma}\left(d\theta^2+\sin^2\theta d\phi^2\right),
\end{equation}
where $M$ is the ADM mass and $0<\gamma\leq 1$. There is a naked singularity at $r=r_s=2M/\gamma$. The JNW naked singularity has a photon sphere for $\gamma>1/2$. The spacetime reduces to Schwarzschild black hole for $\gamma=1$.

\section{Accretion disks around black holes and naked singularities}
\label{sec:disks}
A geometrically thin accretion disk consists of massive particles moving in stable circular timelike geodesics. We consider the Novikov-Thorne model of a thin accretion disk \citep{novikov_1973,page_1974}. Since we are dealing with spherically symmetric, static spacetimes, there are two constants of motions along the timelike geodesics, namely, the specific energy $\tilde{E}$ (energy per unit mass) and the specific angular momentum $\tilde{L}$ of the particles. For a general spherically symmetric, static spacetime of the form
\begin{equation}
ds^2=-A(r)dt^2+B(r)dr^2+C(r)(d\theta^2+\sin^2\theta d\phi^2),
\label{eq:general_metric}
\end{equation}
the timelike geodesic equations in the equatorial plane are given by
\begin{equation}
AB\dot{r}^2+\tilde{V}_{eff}=\tilde{E}^2, \quad \tilde{V}_{eff}=\left(1+\frac{\tilde{L}^2}{C(r)}\right)A(r),
\end{equation}
\begin{equation}
\dot{t}=\frac{\tilde{E}}{A(r)}, \quad \dot{\phi}=\frac{\tilde{L}}{C(r)},
\end{equation}
where an overdot represents differentiation with respect to the affine parameter, and $\tilde{V}_{eff}$ is the effective potential. A stable circular timelike geodesic satisfies $\tilde{V}_{eff}=\tilde{E}^2$, $\tilde{V}'_{eff}=0$ and $\tilde{V}''_{eff}<0$. The first two conditions yield $\tilde{E}$ and $\tilde{L}$ in terms of the circular orbit radius $r$. These are given by \citep{harko_2009}
\begin{equation}
\tilde{E}=\frac{A}{\sqrt{A-C\Omega^2}},\quad \tilde{L}=\frac{C\Omega}{\sqrt{A-C\Omega^2}}, \quad \Omega=\sqrt{\frac{A'}{C'}},
\label{eq:EL}
\end{equation}
where a prime denotes a derivative with respect to $r$, and $\Omega=d\phi/dt$ is the angular momentum of the particles forming the disk. The flux of the electromagnetic radiation emitted from a radial position $r$ of a disk is given by the standard formula \citep{novikov_1973,page_1974}
\begin{equation}
\mathcal{F}(r)=-\frac{\dot{M}}{4\pi\sqrt{-g}}\frac{\Omega'}{(\tilde{E}-\Omega \tilde{L})^2}\int_{r_{in}}^r (\tilde{E}-\Omega \tilde{L})\tilde{L}' dr,
\label{eq:flux}
\end{equation}
where, $r_{in}$ is the inner edge of the disk, $\dot{M}=dM/dt$ is the mass accretion rate and $\sqrt{-g}$ is the four-volume of the induced metric in the equatorial plane.

\subsection{Schwarzschild black hole}
For an accretion disk around a Schwarzschild black hole, the inner edge of the disk is at the innermost stable circular orbit $r_{in}=6M$. This is obtained by solving the circular orbit conditions together with $\tilde{V}''_{eff}=0$. After obtaining the expression for $\tilde{E}$ and $\tilde{L}$ in this case, it is straight forward to show that the flux is given by
\begin{equation}
\mathcal{F}(r)=\frac{3\dot{M}M^{3/2}}{8\pi r^{5/2}(r-3M)}\left[\sqrt{\frac{r}{M}}-\frac{\sqrt{3}}{2}\log\frac{\sqrt{r}-\sqrt{3M}}{\sqrt{r}+\sqrt{3M}}-\sqrt{6}+\frac{\sqrt{3}}{2}\log\frac{\sqrt{2}-1}{\sqrt{2}+1}\right].
\label{eq:sch_flux}
\end{equation}

\subsection{JMN-1 naked singularity}
Using the JMN-1 metric in Eq. (\ref{eq:EL}), We obtain the specific energy and the specific angular momentum of a particle moving in an stable circular orbit in the JMN-1 spacetime. These are, respectively, given by
\begin{equation}
\tilde{E}=\frac{\sqrt{2}(1-M_0)}{\sqrt{2-3M_0}}\left(\frac{r}{R_b}\right)^{M_0/2(1-M_0)},
\end{equation}
\begin{equation}
\tilde{L}=r\sqrt{\frac{M_0}{2-3M_0}}.
\end{equation}
Note that the JMN-1 spacetime does not have any circular orbit for $M_0>2/3$, as both $\tilde{E}$ and $\tilde{L}$ become imaginary. When matched to an exterior Schwarzschild geometry, this singularity model with $M_0>2/3$ has the matching radius $R_b<3M$ and hence, possesses a photon sphere at $r=3M$ which lies in the exterior Schwarzschild geometry \citep{shaikh_2018a}. Therefore, for $M_0>2/3$, we have a single accretion disk lying in the exterior Schwarzschild spacetime and having an inner edge at $r_{in}=6M$. For $M_0<2/3$, the whole interior JMN-1 spacetime possesses circular geodesics in addition to those at radii $r\geq 6M$ in the exterior Schwarszchild spacetime. Note that, for $1/3<M_0<2/3$, the matching radius is $3M<R_b<6M$. Therefore, for $1/3<M_0<2/3$, we have double disks (two disjoint disks)-- an outer disk in the exterior Schwarzschild spacetime and an inner disk in the interior JMN-1 spacetime. The inner edge of the outer disk, which is in the exterior Schwarzschild spacetime, is at the innermost stable orbit $r_{in}=6M$. The particles with angular momentum less than that of the circular orbit at $r_{in}=6M$ plunge in and settle down to circular orbits in the interior JMN-1 spacetime, forming the inner disk in the interior JMN-1. This inner disk extends up to the singularity and has an outer edge at $r=r_{sd}<R_b<6M$ (say). The expression of $r_{sd}$ is obtained by equating the angular momentum $\tilde{L}$ of the innermost stable orbit which is at $r_{in}=6M$ in the exterior Schwarzschild and that of the circular orbit at $r=r_{sd}$ in the interior JMN-1. This gives
\begin{equation}
r_{sd}=R_b\sqrt{3M_0(2-3M_0)}.
\end{equation}
The above expression is valid only for $R_b\leq 6M$, i.e., for $M_0\geq 1/3$. Note that $r_{sd}<R_b$ for $1/3<M_0<2/3$. Therefore, the outer edge of the inner disk is always inside the matching radius. For $M_0=1/3$, $r_{sd}=R_b=6M$, i.e., the outer edge of the inner disk coincides with the inner edge of the outer disk, thereby forming a single continuous disk. Therefore, for $0<M_0\leq 1/3$, we have a single continuous disk extending up to the singularity. The inner part of this disk extends from $r=0$ (singularity) to $r=R_b$ (matching radius) in the interior JMN-1 and the rest is in the exterior Schwarzschild beyond the matching radius $R_b$.  Table \ref{Table1} shows different parameters and the corresponding disk configurations of the JMN-1 naked singularity for different values of $M_0$ which we consider in Sec. \ref{sec:images}.

\begin{table}[h!]
\centering
\caption{Different parameters and the corresponding disk configurations of the JMN-1 naked singularity for different values of $M_0$ which we consider in Sec. \ref{sec:images}.}
 \begin{tabular}{| c | c | c | c | c | c | c |} 
 \hline\hline
 $M_0$ & $R_b$ & Photon & $r_{sd}$& Disk configuration \\
values & (in units of $M$) & sphere & (in units of $M$) & (in units of $M$)  \\
 \hline
  0.70 & 2.857 & Yes & --- & Single disk from $r=r_{in}=6$ to some $r>6$ \\
  0.63 & 3.175 & No & 1.448 & Double disks \\
  & & & & Inner disk: from $r=r_{in}=0$ to $r=r_{sd}=1.448$ \\
  & & & & Outer disk: from $r=r_{in}=6$ to some $r>6$ \\
  0.50 & 4.000 & No & 3.464 & Double disks \\
  & & & & Inner disk: from $r=r_{in}=0$ to $r=r_{sd}=3.464$ \\
  & & & & Outer disk: from $r=r_{in}=6$ to some $r>6$ \\
  0.30 & 6.667 & No & --- & Single disk from $r=r_{in}=0$ to some $r$ \\
 \hline\hline
 \end{tabular}
\label{Table1}
\end{table}

Since for $M_0> 2/3$, we have a single disk with $r_{in}=6M$ in the exterior Schwazschild, the expression for $\mathcal{F}(r)$ in this case is the same as that of the Schwarzchild black hole and is given by Eq. (\ref{eq:sch_flux}). For $1/3<M_0<2/3$, we have double disks and the integration in $\mathcal{F}(r)$ has to be performed separately for each of the disks. For the outer disk, the expression for $\mathcal{F}(r)$ is the same as that of the Schwarzchild black hole. However, for the inner disk, $r_{in}=0$ and $r\leq r_{sd}$. Therefore, for the inner disk, we obtain, after some calculations,
\begin{equation}
\mathcal{F}(r)=\frac{\dot{M}M_0}{4\pi R_b^2}\left(\frac{r}{R_b}\right)^{(3M_0-4)/2(1-M_0)}.
\end{equation}
For $0<M_0\leq 1/3$, we have a single continuous disk extending up to the singularity. In this case, if the emitting point is in the interior JMN-1, i.e., if $0\leq r\leq R_b$, then the corresponding flux is given by the above expression. However, if the emitting point is in the exterior Schwarzschild, i.e., if $r\geq R_b$, then the integration in $\mathcal{F}(r)$ has to be split into two parts-- one from $r=r_{in}=0$ to $r=R_b$ in the interior JMN-1 and other from $r=R_b$ to some $r$ in the exterior Schwarzschild. This gives
\begin{equation}
\mathcal{F}(r)=\frac{\dot{M}M_0}{4\pi R_b^2}+\frac{3\dot{M}M^{3/2}}{8\pi r^{5/2}(r-3M)}\left[\sqrt{\frac{r}{M}}-\frac{\sqrt{3}}{2}\log\frac{\sqrt{r}-\sqrt{3M}}{\sqrt{r}+\sqrt{3M}}-\sqrt{\frac{R_b}{M}}+\frac{\sqrt{3}}{2}\log\frac{\sqrt{R_b}-\sqrt{3M}}{\sqrt{R_b}+\sqrt{3M}}\right].
\end{equation}
Note that $R_b\geq 6M$ in this last case.

\subsection{JNW naked singularity}
The properties of accretion disks in this spacetime have been considered in \cite{chowdhury_2012,gyulchev_2019}. The marginal stable orbits in this spacetime are given by \cite{gyulchev_2019}
\begin{equation}
r_\pm=\frac{3\gamma+1\pm\sqrt{5\gamma^2-1}}{\gamma}.
\end{equation}
For $\gamma>1/2$, we have a photon sphere. In this case, $r_{-}<r_s$ ($r_s$ marks the position of the singularity), and we have a single accretion disk with its inner edge at $r_{in}=r_{+}$. However, for $\gamma<1/2$, we do not have any photon sphere. Similar to the $M_0<2/3$ case of the JMN-1 naked singularity, the $\gamma<1/2$ case of the JNW naked singularity is divided into two subcases. For $1/\sqrt{5}<\gamma<1/2$, $r_s<r_{-}<r_{+}$, and hence, we have double disks. The inner disk extends from the singularity $r=r_s$ to the marginal stable orbit $r=r_{-}$, and outer disk extends from the the marginal stable orbit $r=r_{+}$ to the some outer radius $r>r_+$. This case is similar to the $1/3<M_0<2/3$ case of the JMN-1 naked singularity. For $\gamma=1/\sqrt{5}$, $r_{-}=r_{+}$, thereby forming a single continuous disk. Also, the marginal orbits do not exist for $\gamma<1/\sqrt{5}$ as $r_{\pm}$ becomes imaginary. Therefore, for $\gamma\leq 1/\sqrt{5}$, we have a single continuous disk extending from the singularity $r=r_s$ to some outer radius $r$. This case is similar to the $M_0\leq 1/3$ case of the JMN-1 naked singularity.

 The images of accretion disks have been studied in \cite{gyulchev_2019} for the case when the JNW naked singularity has a photon sphere ($\gamma>1/2$ case). Here, we consider the other ($\gamma<1/2$) case where there are no photon spheres. We numerically integrate Eq. (\ref{eq:flux}) to obtain the flux $\mathcal{F}(r)$ in this JNW naked singularity case.

\section{Tracing the photon geodesics}
\label{sec:tracing}
The emitted photons from the disk undergo gravitational lensing and a fraction of them reach a faraway observer. Therefore, in order to produce the intensity map of the image produces in the observer's sky, we need to solve the null geodesics equations. For a general spherically symmetric, static spacetime given in Eq. (\ref{eq:general_metric}), the null geodesic equations turn out to be
\begin{equation}
\dot{t}=\frac{E}{A(r)}, \quad \dot{\phi}=\frac{L}{C(r)},
\end{equation}
\begin{equation}
\sqrt{AB}\dot{r}=\pm \sqrt{\mathcal{R}(r)}, \quad \mathcal{R}(r)=E^2-(\mathcal{K}+L^2)\frac{A(r)}{C(r)},
\end{equation}
\begin{equation}
C(r)\dot{\theta}=\pm \sqrt{\Theta(\theta)}, \quad \Theta(\theta)=\mathcal{K}-\frac{\cos^2\theta}{\sin^2\theta}L^2,
\end{equation}
where $E$, $L$ and $\mathcal{K}$ are, respectively, the energy, the angular momentum and the Carter constant which remain conserved throughout a photon trajectory. We trace the observed photons backward in time by integrating the above geodesic equations from the observer's position $(r_o,\theta_o)$ to the emission point $(r_e,\pi/2)$ on the accretion disk. However, for the purpose of producing an image of an accretion disk, we have to solve the last two equations only. Combining these two equations and integrating once, we obtain
\begin{equation}
\fint_{\theta_o}^{\pi/2} \frac{d\theta}{\sqrt{\Theta(\theta)}}=\pm \fint_{r_o}^{r_e} \frac{\sqrt{AB}dr}{C\sqrt{\mathcal{R}(r)}},
\label{eq:R-TH}
\end{equation}
where the slash notation $\fint$ indicates that these integrals have to be evaluated along the geodesic, with taking into account all the turning points in the radial or in the polar motion occurring whenever the corresponding potential $\mathcal{R}(r)$ or $\Theta(\theta)$ vanishes. The turning points in $\Theta(\theta)$ are given by
\begin{equation}
\sin^2\theta_{tp}=\frac{L^2}{\mathcal{K}+L^2}
\label{eq:TH_tp}
\end{equation}
which gives following two solutions
\begin{equation}
\theta_{tp1}=\sin^{-1}\frac{|L|}{\sqrt{\mathcal{K}+L^2}}, \quad \theta_{tp2}=\pi-\sin^{-1}\frac{|L|}{\sqrt{\mathcal{K}+L^2}},
\end{equation}
where $0\leq \theta_{tp1}\leq \pi/2$ and $\pi/2\leq \theta_{tp2}\leq \pi$, i.e., $\theta_{tp1}$ and $\theta_{tp2}$ lie, respectively, in the upper and lower half about the equatorial plane. The photons which do not take turn in $\theta$, the left hand side of (\ref{eq:R-TH}) becomes
\begin{equation}
\fint_{\theta_o}^{\pi/2} \frac{d\theta}{\sqrt{\Theta(\theta)}}=\int_{\theta_o}^{\pi/2} \frac{d\theta}{\sqrt{\Theta(\theta)}}
\label{eq:TH-0}
\end{equation}
However, photons which undergo strong gravitational lensing may undergo many windings around the central object and hence, may encounter many turnings at the turning points $\theta_{tp1}$ and $\theta_{tp2}$. Note that $\theta_{tp1}\leq \theta_o\leq \theta_{tp2}$. Since we are integrating the geodesics backward in time from the observer's position to the emitting point on the disk, i.e., we are shooting photons from the observer's position towards the central objects at different impact parameters, then depending on the initial direction of a photon, i.e., depending on the sign on the right hand side of Eq. (\ref{eq:R-TH}) at the observer's position, the photon first encounters one of the above turning points. Let us first choose the `$+$' sign in (\ref{eq:R-TH}), i.e., we choose a photon with $d\theta/dr>0$ initially at the observer's position. Therefore, since the radial coordinate decreases initially along the photon geodesic which we shoot from the observer's position, the coordinate $\theta$ has to decrease initially. Then, if the photon encounter any turning point in $\theta$, it does so at $\theta_{tp1}$ first and then at $\theta_{tp2}$. After that, it may repeat the same and undergoes multiple turnings at both the turning points, depending on its amount of deflection. If the photon hit the emitting point on the disk after encountering its last turn at $\theta_{tp1}$, then the left hand side of (\ref{eq:R-TH}) becomes
\begin{eqnarray}
\fint_{\theta_o}^{\pi/2} \frac{d\theta}{\sqrt{\Theta(\theta)}}&=&\int_{\theta_o}^{\theta_{tp1}} \frac{d\theta}{\sqrt{\Theta(\theta)}}-\int_{\theta_{tp1}}^{\theta_{tp2}} \frac{d\theta}{\sqrt{\Theta(\theta)}}+\int_{\theta_{tp2}}^{\theta_{tp1}} \frac{d\theta}{\sqrt{\Theta(\theta)}}-\cdots \nonumber \\
& & \cdots +\int_{\theta_{tp2}}^{\theta_{tp1}} \frac{d\theta}{\sqrt{\Theta(\theta)}} -\int_{\theta_{tp1}}^{\pi/2} \frac{d\theta}{\sqrt{\Theta(\theta)}} \nonumber \\
&=& \int_{\theta_o}^{\theta_{tp1}} \frac{d\theta}{\sqrt{\Theta(\theta)}}-n_1\int_{\theta_{tp1}}^{\theta_{tp2}} \frac{d\theta}{\sqrt{\Theta(\theta)}}-\int_{\theta_{tp1}}^{\pi/2} \frac{d\theta}{\sqrt{\Theta(\theta)}},
\label{eq:TH-1}
\end{eqnarray}
where we have changed the sign of $d\theta$ after each turning, and $n_1(=0,2,4...)$ is the total number of trajectory arcs covered by the photon between $\theta_{tp1}$ and $\theta_{tp2}$ and between $\theta_{tp2}$ and $\theta_{tp1}$. Note that $n_1$ must be zero or even integer in order that the photon hit the emitting point after taking its last turn at $\theta_{tp1}$. However, if the photon hit the emitting point after encountering its last turn at $\theta_{tp2}$, then we obtain
\begin{eqnarray}
\fint_{\theta_o}^{\pi/2} \frac{d\theta}{\sqrt{\Theta(\theta)}}&=&\int_{\theta_o}^{\theta_{tp1}} \frac{d\theta}{\sqrt{\Theta(\theta)}}-\int_{\theta_{tp1}}^{\theta_{tp2}} \frac{d\theta}{\sqrt{\Theta(\theta)}}+\int_{\theta_{tp2}}^{\theta_{tp1}} \frac{d\theta}{\sqrt{\Theta(\theta)}}-\cdots \nonumber \\
& & \cdots -\int_{\theta_{tp1}}^{\theta_{tp2}} \frac{d\theta}{\sqrt{\Theta(\theta)}} +\int_{\theta_{tp2}}^{\pi/2} \frac{d\theta}{\sqrt{\Theta(\theta)}} \nonumber \\
&=& \int_{\theta_o}^{\theta_{tp1}} \frac{d\theta}{\sqrt{\Theta(\theta)}}-n_2\int_{\theta_{tp1}}^{\theta_{tp2}} \frac{d\theta}{\sqrt{\Theta(\theta)}}+\int_{\theta_{tp2}}^{\pi/2} \frac{d\theta}{\sqrt{\Theta(\theta)}},
\label{eq:TH-2}
\end{eqnarray}
where, $n_2=1,3,5...$. Note that $n_2$ must be odd integer in order that the photon hit the emitting point after taking its last turn at $\theta_{tp2}$. Note also that $\theta_{tp2}=\pi-\theta_{tp1}$ and $\Theta(\pi-\theta)=\Theta(\theta)$ as $0\leq\theta\leq \pi$. Therefore, if we use the transformation $\theta\rightarrow \pi-\theta$, then the last integral on the right hand side of the last equation becomes
\begin{equation}
\int_{\theta_{tp2}}^{\pi/2} \frac{d\theta}{\sqrt{\Theta(\theta)}}=-\int_{\theta_{tp1}}^{\pi/2} \frac{d\theta}{\sqrt{\Theta(\theta)}}.
\end{equation}
Therefore, Eqs. (\ref{eq:TH-1}) and (\ref{eq:TH-2}) can be combined into a single one given by
\begin{equation}
\fint_{\theta_o}^{\pi/2} \frac{d\theta}{\sqrt{\Theta(\theta)}}=\int_{\theta_o}^{\theta_{tp1}} \frac{d\theta}{\sqrt{\Theta(\theta)}}-n\int_{\theta_{tp1}}^{\theta_{tp2}} \frac{d\theta}{\sqrt{\Theta(\theta)}}-\int_{\theta_{tp1}}^{\pi/2} \frac{d\theta}{\sqrt{\Theta(\theta)}},
\label{eq:TH-3}
\end{equation}
where $n=0,1,2,3...$.

Similarly, we now choose the `$-$' sign in (\ref{eq:R-TH}), i.e., we choose a photon with $d\theta/dr<0$ initially at the observer's position. Therefore, since the radial coordinate decreases initially along the photon geodesic which we shoot from the observer's position, the coordinate $\theta$ has to increase initially. Then, if the photon encounter any turning point in $\theta$, it does so at $\theta_{tp2}$ first and then at $\theta_{tp1}$. After that, it may repeat the same and undergoes multiple turnings at both the turning point. Following the same procedure discussed above, for such a photon, we obtain
\begin{equation}
\fint_{\theta_o}^{\pi/2} \frac{d\theta}{\sqrt{\Theta(\theta)}}=\int_{\theta_o}^{\theta_{tp2}} \frac{d\theta}{\sqrt{\Theta(\theta)}}-m\int_{\theta_{tp2}}^{\theta_{tp1}} \frac{d\theta}{\sqrt{\Theta(\theta)}}-\int_{\theta_{tp2}}^{\pi/2} \frac{d\theta}{\sqrt{\Theta(\theta)}},
\label{eq:TH-4}
\end{equation}
where $m=0,1,2,3...$.

We now consider the radial integral. Let $r_{tp}$ be the outermost turning point in $r$ such that $\mathcal{R}(r_{tp})=0$. If the photon hit the emitting point on the disk before it reaches the turning point $r_{tp}$, then the radial integral in (\ref{eq:R-TH}) can be written as
\begin{equation}
\fint_{r_o}^{r_e} \frac{\sqrt{AB}dr}{C\sqrt{\mathcal{R}(r)}}=\int_{r_o}^{r_e} \frac{\sqrt{AB}dr}{C\sqrt{\mathcal{R}(r)}}.
\label{eq:R-1}
\end{equation}
However, if it hits the disk after encountering the turning point, then we have
\begin{equation}
\fint_{r_o}^{r_e} \frac{\sqrt{AB}dr}{C\sqrt{\mathcal{R}(r)}}=\int_{r_o}^{r_{tp}} \frac{\sqrt{AB}dr}{C\sqrt{\mathcal{R}(r)}}-\int_{r_{tp}}^{r_e} \frac{\sqrt{AB}dr}{C\sqrt{\mathcal{R}(r)}}.
\label{eq:R-2}
\end{equation}

The apparent shape of an image is obtained by using the celestial coordinates $\alpha$ and $\beta$ which lie in the celestial plane perpendicular to the line joining the observer and the center of the spacetime geometry. The coordinates $\alpha$ and $\beta$ are defined by \citep{celestial}
\begin{equation}
\alpha=\lim_{r_o\to\infty}\left(-r^2\sin\theta\frac{d\phi}{dr}\Big\vert_{(r_o,\theta_o)}\right),
\end{equation}
\begin{equation}
\beta=\lim_{r_o\to\infty}\left(r^2\frac{d\theta}{dr}\Big\vert_{(r_o,\theta_o)}\right),
\end{equation}
where $(r_o,\theta_o)$ are the position coordinates of a distant observer. After using the geodesic equations and taking the limit, we obtain
\begin{equation}
\alpha=-\frac{\xi}{\sin\theta_0},
\label{eq:alpha}
\end{equation}
\begin{equation}
\beta=\pm \sqrt{\eta-\xi^2\cot^2\theta_0},
\label{eq:beta}
\end{equation}
where the $\pm$ sign in the last equation is the same as that on the right hand side of (\ref{eq:R-TH}), $\xi=L/E$ and $\eta=\mathcal{K}/E^2$.

Now, the $\theta$-integration can be performed analytically by putting $\cos\theta=x$. We obtain up to a integration constant
\begin{equation}
\int \frac{d\theta}{\sqrt{\Theta(\theta)}}=-\frac{1}{\sqrt{\mathcal{K}+L^2}}\sin^{-1}\left[\sqrt{\frac{\mathcal{K}+L^2}{\mathcal{K}}}\cos\theta\right].
\end{equation}
Note that $0\leq \theta_{tp1}\leq \pi/2$ and $\pi/2\leq \theta_{tp2}\leq \pi$. Therefore, from (\ref{eq:TH_tp}), we obtain $\cos\theta_{tp1}=\sqrt{\frac{\mathcal{K}}{\mathcal{K}+L^2}}$ and $\cos\theta_{tp2}=-\sqrt{\frac{\mathcal{K}}{\mathcal{K}+L^2}}$. Using these and the last equation in Eqs. (\ref{eq:TH-0}), (\ref{eq:TH-3}) and (\ref{eq:TH-4}), Eq. (\ref{eq:R-TH}) can be rewritten as
\begin{equation}
\fint_{r_o}^{r_e} \frac{\sqrt{AB}dr}{C\sqrt{\mathcal{R}(r)/E^2}}=-\frac{1}{\sqrt{\eta+\xi^2}}\left[k\pi-sign(\beta)\sin^{-1}\left(\sqrt{\frac{\eta+\xi^2}{\eta}}\cos\theta_o\right)\right],
\end{equation}
where $sign(\beta)$ is the sign of the $\beta$-coordinates given in (\ref{eq:beta}), i.e., the sign on the right hand side of (\ref{eq:R-TH}), and the radial part on the left hand side of the above equation is given by Eq. (\ref{eq:R-1}) and Eq. (\ref{eq:R-2}), respectively, when the photon hits the emitting point before and after encountering the turning point. Here, we have multiplied both sides of the above equation by $E$ and replaced $n$ and $m$ by a common integer $k(=0,1,2,3...)$.

We now discuss the procedure to find out the solution $r_e$ of the above equation such that $r_e$ lies within the inner and outer boundaries of an accretion disk. We first set the observer's position $(r_o,\theta_o)$ and the coordinates $(\alpha,\beta)$ in the observer sky. This fixes the impact parameter $\xi$ and $\eta$, i.e., $L$ and $\mathcal{K}$. We then increase $k$ from $0$ in steps of $1$ until a root $r_e$ of the above equation is found. If the root $r_e$ is found for a given $k$ and it lies outside a disk, we ignore that root and proceed further by increasing $k$. We increase $k$ up to a maximum value of $k_{max}=15$. Note that, while finding $r_e$, we first consider Eq. (\ref{eq:R-1}) for the radial part and increase $k$ from $0$ to $k_{max}$. If a root is found, we stop there. If not, then we repeat the same by considering Eq. (\ref{eq:R-2}) for the radial part and increase $k$ from $0$ to $k_{max}$.

The range of $k$ from $k=0$ to $k_{max}=15$ covers all the photons which undergo bending roughly up to $15\pi$. For black holes, $k_{max}=3$ or $4$ is sufficient as strong deflection of light takes place for impact parameters very close to the critical impact parameter corresponding to the photon sphere. The deflection angle diverges logarithmically as the impact parameter approaches the critical value. For example, for Schwarzschild black hole, the deflection angle is infinite (theoretically) at the critical impact parameter $3\sqrt{3}\simeq 5.19615$ (in units of $M$) and is $3\pi$ (say) at the impact parameter $5.19640$. Therefore, if we choose $k_{max}=3$, i.e., if we only consider photons with deflection roughly up to $3\pi$ and ignore those having deflection more than $3\pi$, then this means that we are ignoring photons within the impact parameter gap $(5.19640-5.19615)=0.00025$ in the observers sky. Ignoring this small gap does not affect the images. However, for the naked singularity models without any photon sphere considered in the next section ($M_0<2/3$ and $\gamma<1/2$ cases in the next section), the deflection angle can still be large and is not logarithmically divergent unlike black hole case. In such cases, the gap between two impact parameters which have $2\pi$ difference in their deflection angles is not small as we do not have the logarithmic divergence of the bending angle. Hence, we have to choose $k_{max}$ carefully in such cases. We have found that, for the set of parameter values considered in such cases, the maximum deflection is roughly $10\pi$ for $M_0=0.63$ JMN-1 model. For $\gamma<1/2$ JNW models or for other lower values of $M_0$ of JMN-1 model, the maximum deflection is even lower. Therefore, it would have been sufficient to choose $k_{max}=10$ or $11$ in such cases. However, we have chosen $k_{max}=15$ during our numerical calculations. Therefore, in such cases of the naked singularity without a photon sphere, we have taken all the photons into consideration.

The photon flux as detected by a distant observer is given by \citep{bambi_2012}
\begin{equation}
F_{obs}(r)=g^4\mathcal{F}(r),\quad g = \frac{k_\alpha u^\alpha_{o}}{k_\beta u^\beta_{e}}=\frac{\sqrt{A-C\Omega^2}}{1-\xi\Omega},
\end{equation}
where $\mathcal{F}(r)$ is the flux obtained in the previous section, $g$ is the redshift factor, $u^\mu_{o} = (1,0,0,0)$ is the four-velocity of the distant observer (who is at infinity), $u^\mu_{e} = (\dot{t},0,\dot{\phi},0)$ is the four-velocity of the timelike geodesic at the emitting point on the accretion disk, and $k^\mu$ is the four-velocity of the photons obtained in the previous section. Therefore, once we find $r_e$ for a given set of value of $(\alpha,\beta)$, we assign to it the above observed redshifted flux value by putting $r=r_e$ into it.

\section{Images of black holes and naked singularities with accretion disks}
\label{sec:images}

\begin{figure}[ht]
\centering
\subfigure[~$M = 1.0$, Schwarzschild black hole]{\includegraphics[scale=0.53]{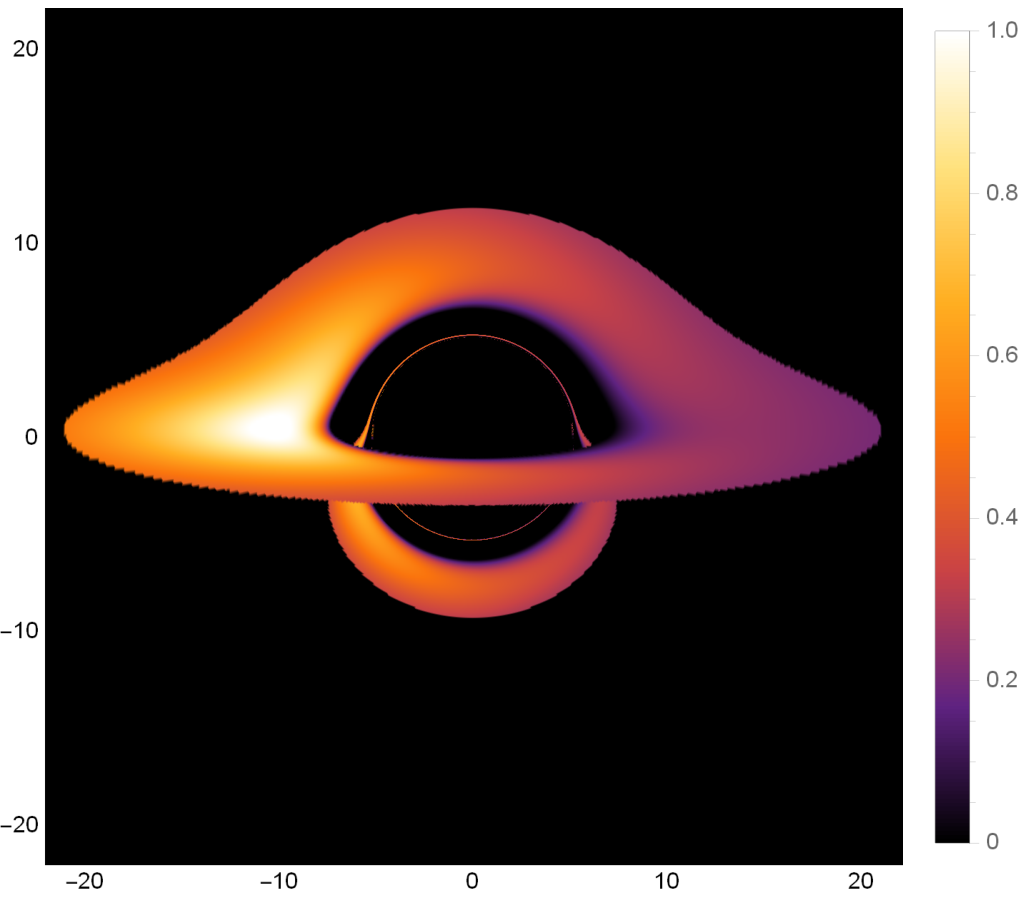}\label{fig:JMN1a}}\hspace{0.1cm}
\subfigure[~$M_0 = 0.7$, JMN-1 naked singularity]{\includegraphics[scale=0.56]{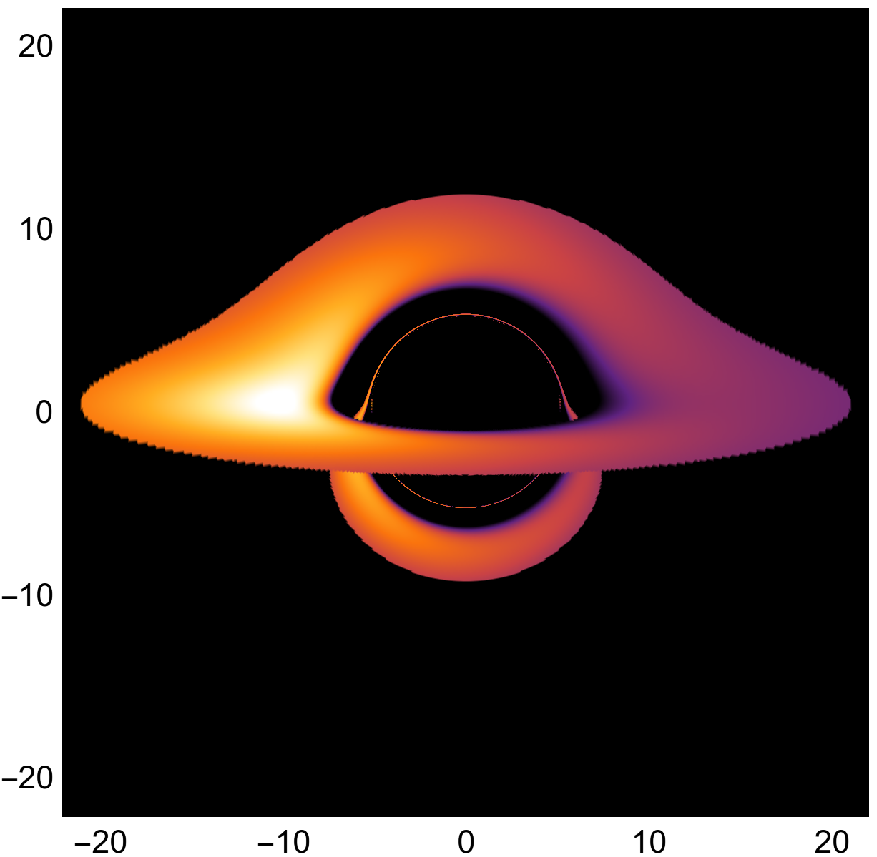}\label{fig:JMN1b}}\hspace{0.1cm}
\subfigure[~$M_0 = 0.63$, JMN-1 naked singularity]{\includegraphics[scale=0.56]{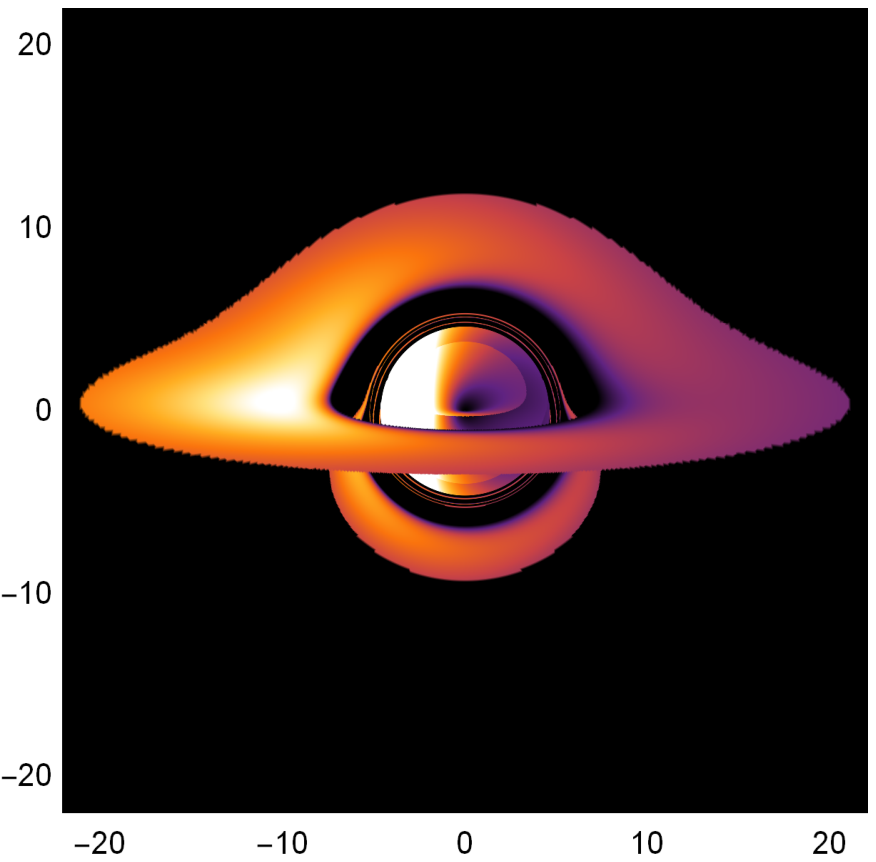}\label{fig:JMN1c}}
\subfigure[~$M_0 = 0.50$, JMN-1 naked singularity]{\includegraphics[scale=0.59]{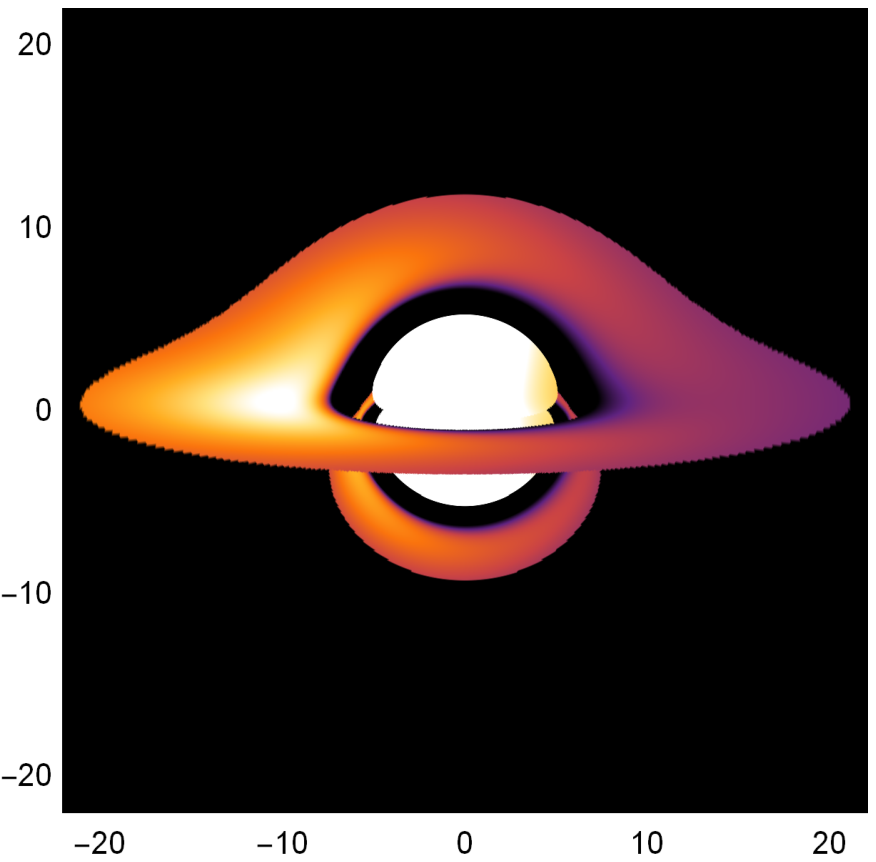}\label{fig:JMN1d}}\hspace{0.1cm}
\subfigure[~$M_0 = 0.30$, JMN-1 naked singularity]{\includegraphics[scale=0.59]{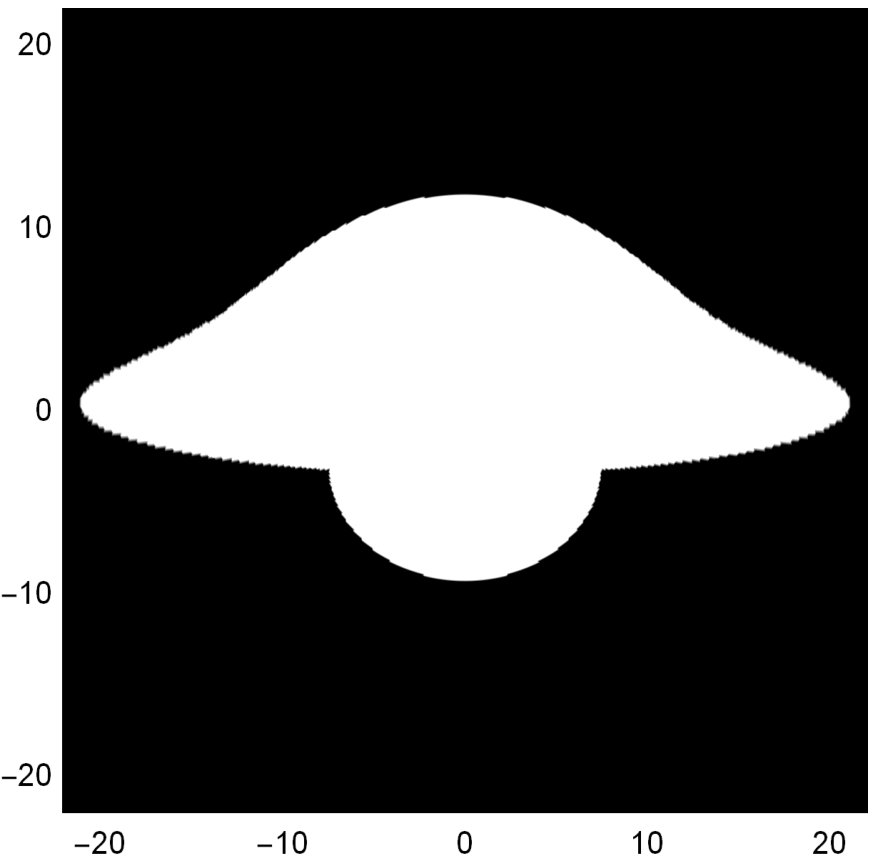}\label{fig:JMN1e}}\hspace{0.1cm}
\subfigure[~Zoom in version of (c)]{\includegraphics[scale=0.58]{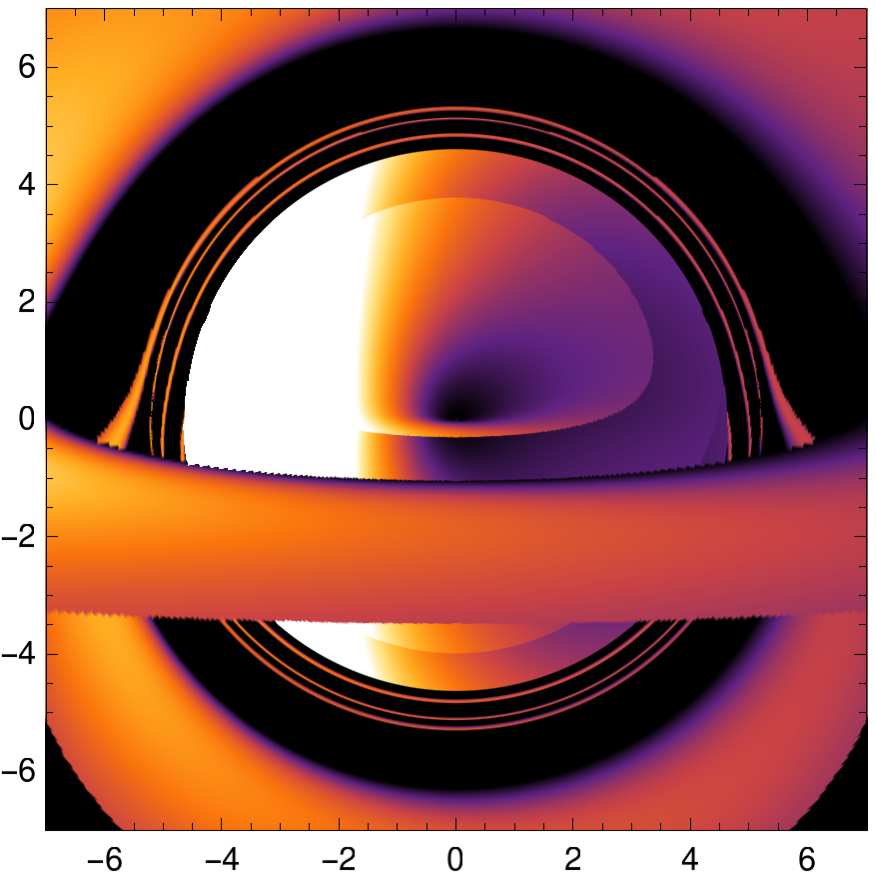}\label{fig:JMN1f}}
\caption{The images of a Schwarzschild black hole and a JMN-1 naked singularity with accretion disks [(a)-(f)]. For $M_0\geq 2/3$, the JMN-1 naked singularity has a photon sphere and has a single accretion disk with an inner edge lying outside the photon sphere [(b)]. For $M_0<2/3$, it does not have any photon sphere and has two accretion disks for $2/3<M_0<1/3$ [(c) and (d)] and a single disk extending up to the singularity for $M_0\leq 1/3$ [(e)]. The disk configurations corresponding to different $M_0$ values are shown in Table \ref{Table1}. The outer edge of the outer disk is at $r=20M$, and the observer's inclination angle is $\theta_{o}=80^{\circ}$. The observer is placed at the radial coordinate $r=10^4M$, which corresponds effectively to the asymptotic infinity. In order to get rid of the parameters $M$ and $\dot{M}$, we have normalized the fluxes by the maximum flux observed for the Schwarzschild black hole. Also, we have plotted the square-root of the normalized flux for better looking. All spatial coordinates are in units of $M$.}
\label{fig:JMN1}
\end{figure}

\begin{figure}[ht]
\centering
\subfigure[~$\gamma=1$, Schwarzschild black hole]{\includegraphics[scale=0.6]{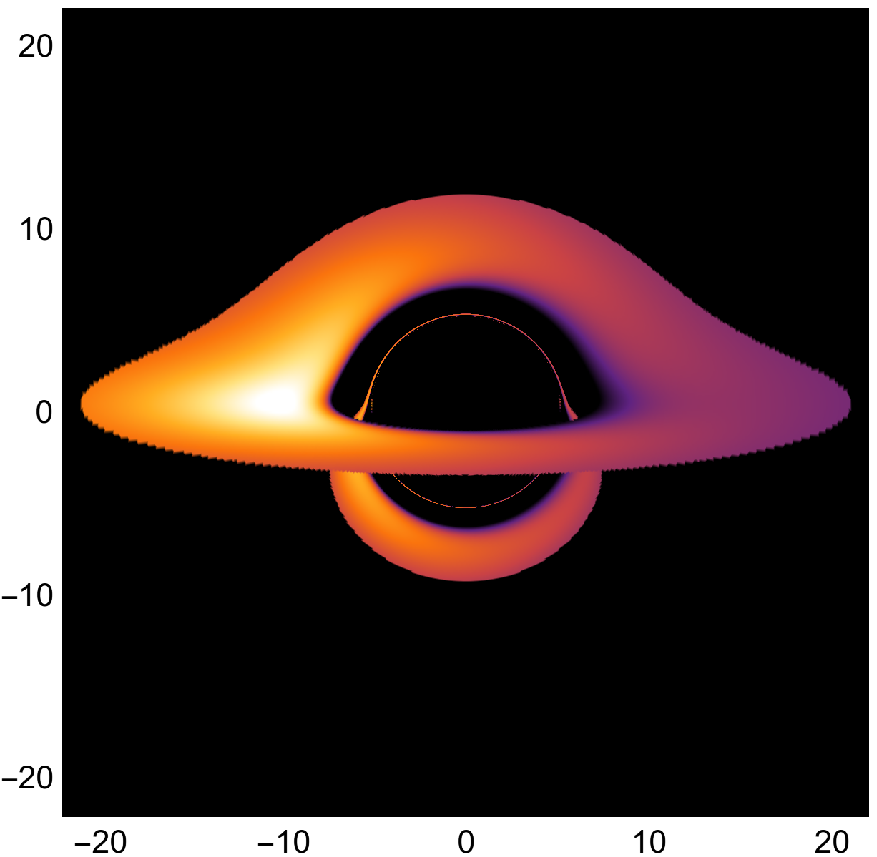}\label{fig:JNWa}}\hspace{0.1cm}
\subfigure[~$\gamma = 0.75$, JNW naked singularity]{\includegraphics[scale=0.6]{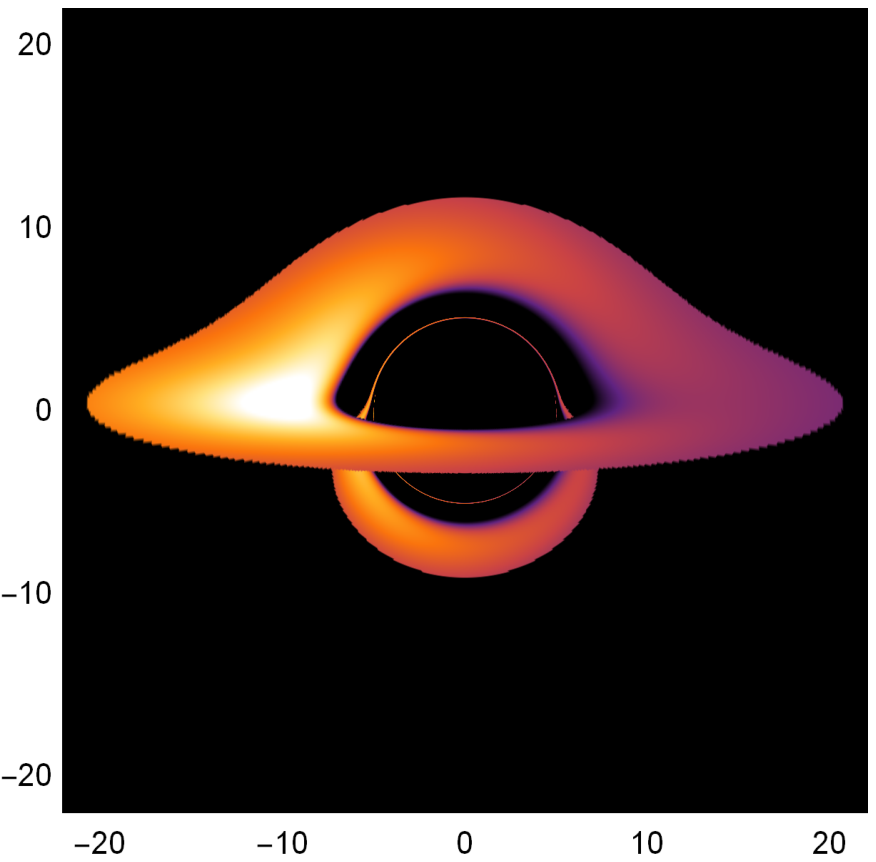}\label{fig:JNWb}}\hspace{0.1cm}
\subfigure[~$\gamma = 0.51$, JNW naked singularity]{\includegraphics[scale=0.6]{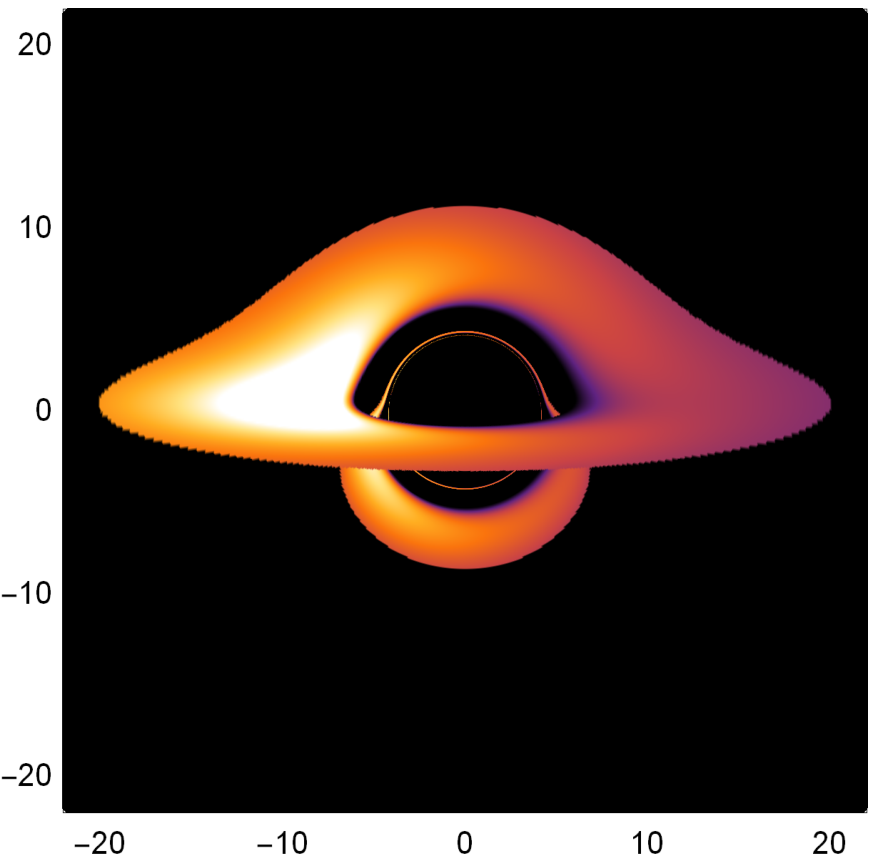}\label{fig:JNWc}}
\subfigure[~$\gamma = 0.45$, JNW naked singularity]{\includegraphics[scale=0.6]{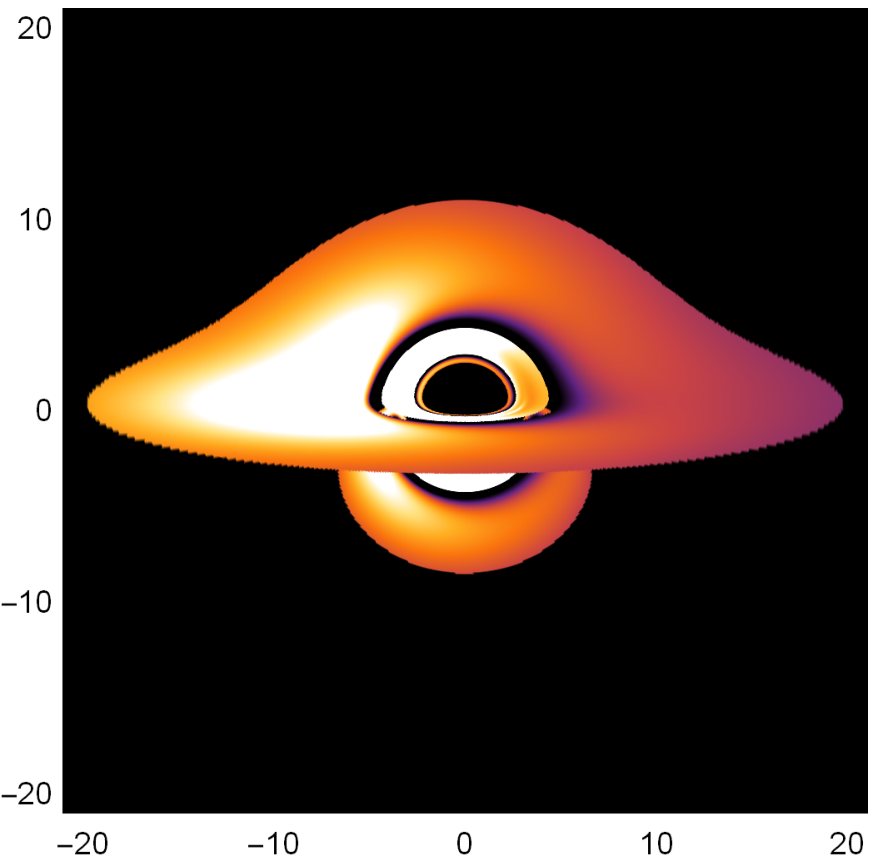}\label{fig:JNWd}}\hspace{0.1cm}
\subfigure[~$\gamma = 0.30$, JNW naked singularity]{\includegraphics[scale=0.6]{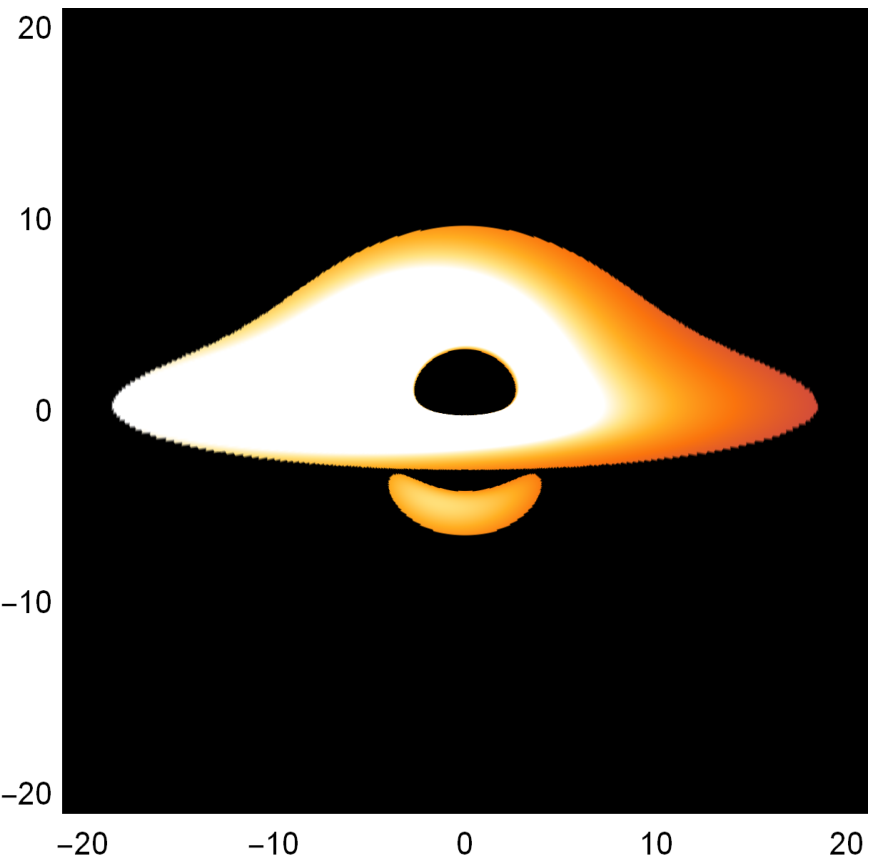}\label{fig:JNWe}}\hspace{0.1cm}
\subfigure[~Zoom in version of (d)]{\includegraphics[scale=0.59]{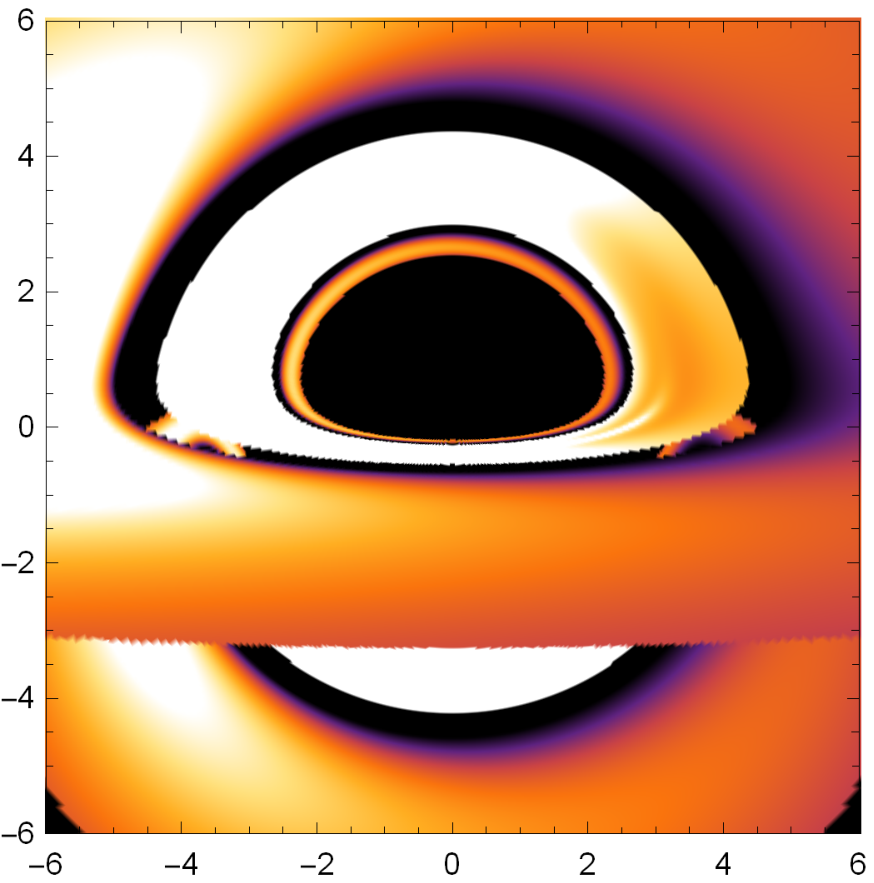}\label{fig:JNWf}}
\caption{The images of a Schwarzschild black hole and a JNW naked singularity with accretion disks [(a)-(f)]. For $\gamma\geq 1/2$, the JNW naked singularity has a photon sphere and has a single accretion disk with an inner edge lying outside the photon sphere [(b) and (c)]. For $\gamma<1/2$, it does not have any photon sphere and has two accretion disks for $1/2<\gamma<1/\sqrt{5}$ [(d)] and a single disk extending up to the singularity for $\gamma\leq 1/\sqrt{5}$ [(e)]. The outer edge of the outer disk is at $r=20M$, and the observer's inclination angle is $\theta_{o}=80^{\circ}$. The observer is placed at the radial coordinate $r=10^4M$, which corresponds effectively to the asymptotic infinity. In order to get rid of the parameters $M$ and $\dot{M}$, we have normalized the fluxes by the maximum flux observed for the Schwarzschild black hole. Also, we have plotted the square-root of the normalized flux for better looking. All spatial coordinates are in units of $M$.}
\label{fig:JNW}
\end{figure}

The images of the JMN-1 and JNW naked singularities for the different cases are, respectively, shown in Fig. \ref{fig:JMN1} and Fig. \ref{fig:JNW}. Note that, when the JMN-1 naked singularity has a photon sphere, its image [Fig. \ref{fig:JMN1b}] mimic that of the Schwarzschild black hole [Fig. \ref{fig:JMN1a}] both qualitatively and quantitatively, as the accretion disk around the JMN-1 singularity in such case lies in the exterior Schwarzschild geometry, and hence, both the flux distribution and the corresponding image are the same as those of the Schwarzschild black hole. Therefore, it is difficult to distinguish a JMN-1 singularity having a photon sphere from a black hole. Similar conclusion was drawn when images in the presence of a spherical accretion were considered \citep{shaikh_2018a}. However, when the JNW naked singularity has a photon sphere, its images [Figs. \ref{fig:JNWb} and \ref{fig:JNWc}] mimic that of the Schwarzschild black hole [Fig. \ref{fig:JNWa}] qualitatively and differ quantitatively. This is because, as we decrease $\gamma$ in such case, the brightness of the image increases because of the increase in the emitted flux. See \cite{gyulchev_2019} for a detailed study of this JNW case.

On the other hand, in the absence of a photon sphere (which is the case when $M_0<2/3$ for JMN-1 or $\gamma<1/2$ for JNW), the images of both class of the naked singularities [Figs. \ref{fig:JMN1c}--\ref{fig:JMN1f} and \ref{fig:JNWd}--\ref{fig:JNWf}] significantly differ from that of the black hole. In such case, extra images appear around the naked singularities. Therefore, such singularity models can be clearly distinguished from a black hole. Note that the image of the inner disk for $M_0=0.5$ JMN-1 model [Figs. \ref{fig:JMN1d}] or that of the single disk extending up to the singularity for $M_0=0.3$ JMN-1 model [Figs. \ref{fig:JMN1e}] are much brighter as compared to the image of the black hole. The reason for this is that the flux of the emission emitted from the vicinity of the JMN-1 naked singularity in such cases is much higher than that from a disk around the Schwarzschild black hole. See \cite{JMN2} for a similar case of higher flux (from a disk around a naked singualarity) as compared to that of a black hole. However, a part of the images of the inner disk for $M_0=0.63$ JMN-1 model [Figs. \ref{fig:JMN1c}] appears much less bright because of the high redshift. For the JNW models also, as we decrease $\gamma$, the brightness of the image increases because of the increase in the emitted flux [Figs. \ref{fig:JNWd} and \ref{fig:JNWe}].

\begin{figure}[ht]
\centering
\subfigure[~$M_0=0.3$ (blue), $\gamma=0.3$ (red)]{\includegraphics[scale=0.73]{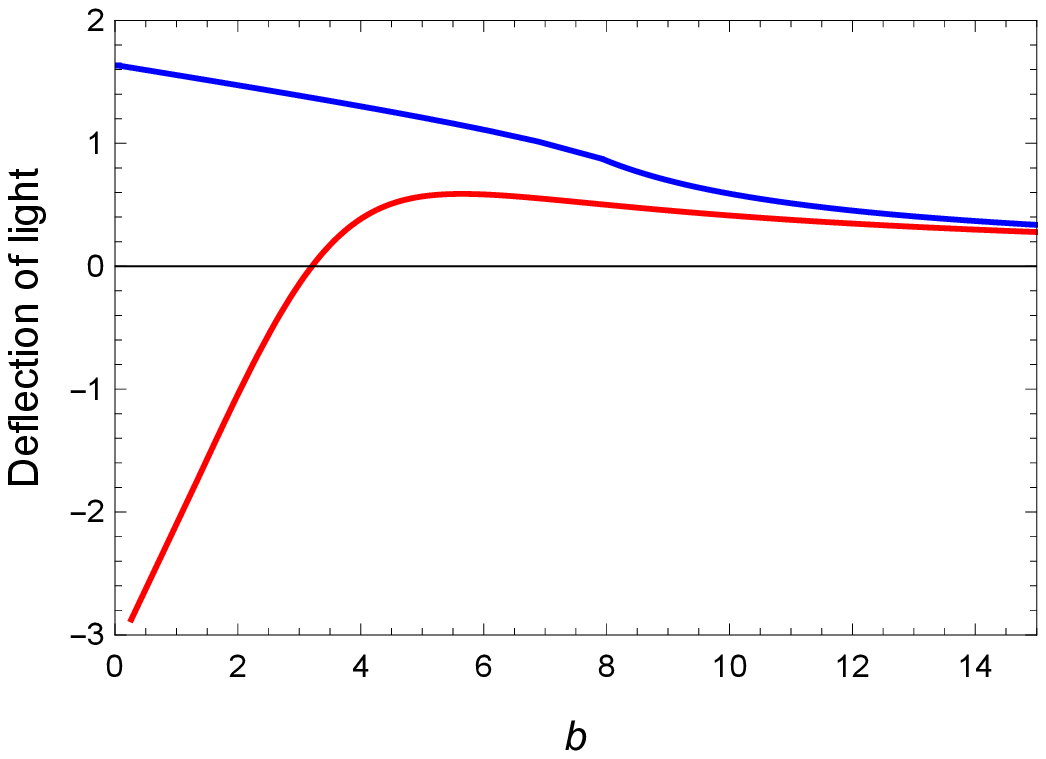}\label{fig:deflection}}\hspace{0.1cm}
\subfigure[~$M_0=0.3$ (blue), $\gamma=0.3$ (red)]{\includegraphics[scale=0.6]{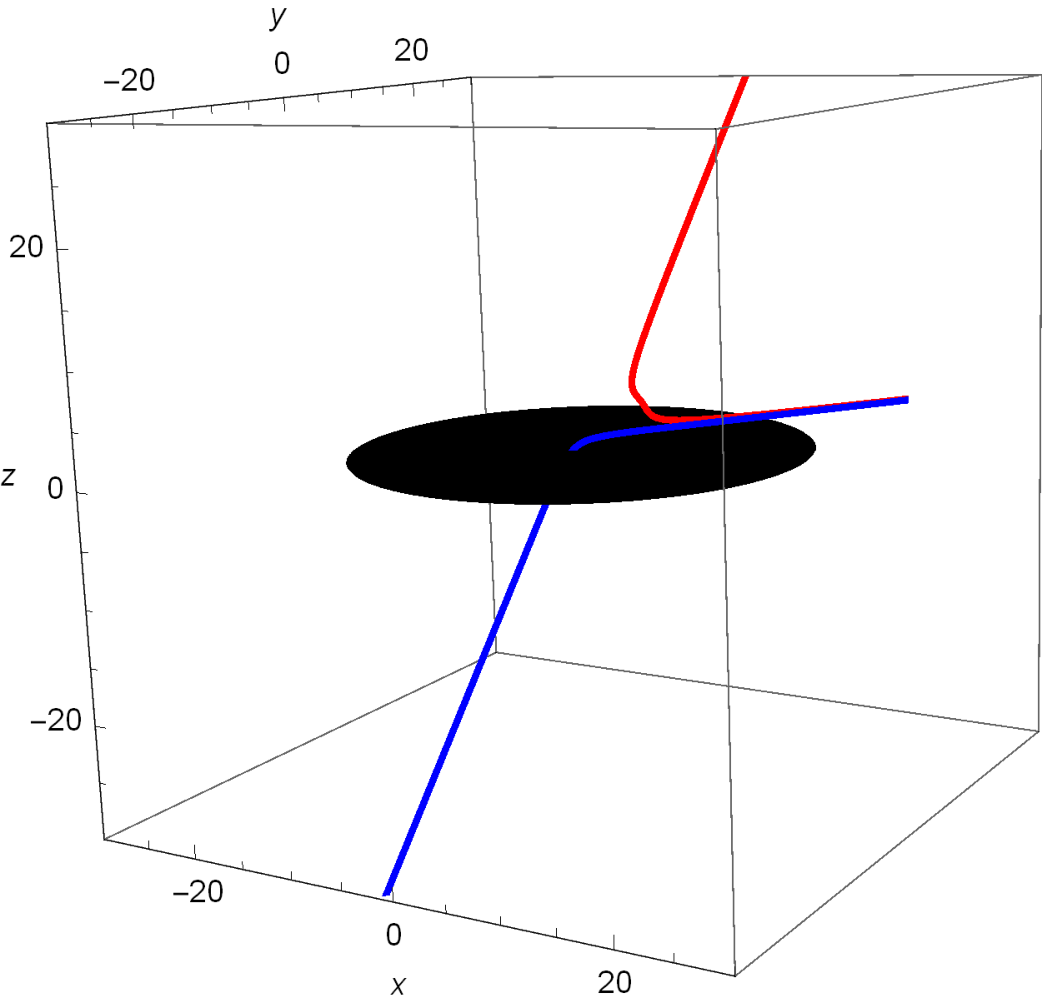}\label{fig:geodesic}}
\caption{(a) Deflection angle of light as a function of the impact parameter $b=L/E$ in the equatorial plane of the JMN-1 naked singularity with $M_0=0.3$ (blue) and the JNW naked singularity with $\gamma=0.3$ (red). (b) Photon geodesic in the JMN-1 naked singularity with $M_0=0.3$ (blue) and the JNW naked singularity with $\gamma=0.3$ (red). Both the geodesics have the same impact parameter given by the celestial coordinates $(\alpha,\beta)=(1.0,1.0)$. The geodesics are originated from the observer's location $(r_o,\theta_o,\phi_o)=(10^4M,80^{\circ},0)$. The black disk shows the accretion disk.}
\label{fig:deflection_geodesic}
\end{figure}

Note that, in the absence of a photon sphere, the images of both class of the naked singularity not only differ from that of the black hole but also from those of one another. For example, in such cases, the image of the JNW singularity [Figs. \ref{fig:JNWd} and \ref{fig:JNWe}] contains a hole (which is devoid of flux) at its center whereas that of the JMN-1 singularity [Figs. \ref{fig:JMN1c}--\ref{fig:JMN1e}] does not. This difference can be explained from the deflection of light passing very close to the singularities. Figure \ref{fig:deflection} shows the deflection angle of light in the equatorial plane. Note that, for small impact parameter, the deflection is positive in JMN-1 spacetime and negative in JNW spacetime. Therefore, a light ray when passes very close to the JMN-1 singularity undergoes positive deflection whereas it undergoes negative deflection when it passes very close to the JNW naked singularity. In Fig. \ref{fig:geodesic}, we  have shown the geodesics of a photon which is shoot from the observer's position towards the singularities with small impact parameter. Note that, because of the positive deflection in the JMN-1 naked singularity case, the geodesic bends towards the singularity and hits the accretion disk, thereby indicating that the photon gets emitted from this hitting point and reaches the observer. On the other hand, because of the negative deflection in the JNW naked singularity case, the geodesic bends away from the singularity and miss the accretion disk, thereby indicating that the photon, if emitted from the disk with the corresponding impact parameter, does not reach the observer. This gives a black spot at the corresponding celestial coordinates. This explains why the images of the JNW naked singularity with $\gamma<1/2$ contain holes at their center even though there exist disks around or up to the singularity.

\section{conclusions}
\label{sec:conclusions}
In this work, we have studied the images of thin accretion disks around two classes of naked singularities, namely the JMN-1 and the JNW naked singularity, and compared them with those of Schwarzschild black hole. It turns out that, when the JMN-1 naked singularity has a photon sphere, it is difficult to distinguish it from a Scwarzschild black hole as its images mimic that of the black hole both qualitatively and quantitatively. However, for the JNW naked singularity with a photon sphere, the images mimic that of a black hole qualitatively, although there is a quantitative difference (see also \cite{gyulchev_2019} for this JNW case). It, therefore, follows that further and more detailed analysis of the images and shadows structure in such cases is needed to confirm or otherwise the existence of an event horizon for the compact objects such as the galactic centers.

On the other hand, in the absence of a photon sphere, the images of both class of the naked singularities significantly differ from that of the black hole. In such case, the naked singularities can be clearly distinguished from a black hole through the observation of their images. Moreover, the images of both class of the naked singularities in this case also differ from one another, thereby allowing them to be distinguish from one another through the observation of the images.

\end{document}